\documentclass{appolb}

\begin{document}
\pagestyle{plain}
\newcount\eLiNe\eLiNe=\inputlineno\advance\eLiNe by -1
\title{MAGNETIC ORDER IN TRANSITION METAL OXIDES \\ 
       WITH ORBITAL DEGREES OF FREEDOM 
}
\author{Andrzej M. Ole\'s
\address{Institute of Physics, Jagellonian University,\\
Reymonta 4, PL-30059 Krak\'ow, Poland\\
e-mail: amoles@if.uj.edu.pl
}}
\date{\it 21 June 2001}
\maketitle

\begin{abstract}
We investigate the frustrated magnetic interactions in cubic 
transition metal oxides with orbital degeneracy. The $e_g$ orbitals
order easier and their ordering explains the $A$-type antiferromagnetic 
phase in KCuF$_3$ and LaMnO$_3$. In $t_{2g}$ systems the magnetic 
order changes at a transition from an orbital liquid to orbital ordered 
states. The fluctuations of $t_{2g}$ orbitals play a prominent role in 
LaVO$_3$ and YVO$_3$, where they compete with the Jahn-Teller effect and 
trigger the C-type antiferromagnetic order.
\end{abstract}


\section{ Spin-orbital physics in transition metal oxides }
\label{sec:intro}

Large on-site Coulomb interactions $\propto U$ in transition metal oxides 
suppress charge fluctuations and lead to the (partial) localization of 
$d$ electrons which interact by the effective superexchange interactions.
When such localized electrons occupy degenerate orbital states, one 
has to consider orbital degrees of freedom at equal footing with electron 
spins \cite{Tok00}. The importance of the orbital degrees of freedom in 
such systems has been emphasized long ago for cuprates \cite{Kug82} and 
for V$_2$O$_3$ \cite{Cas78}, when it was also realized that ferromagnetic 
(FM) superexchange could be induced by the Hund's exchange interaction 
$\propto J_H$ \cite{Cla75}, but only recently it has been fully 
appreciated that {\it the orbital physics\/} leads to several novel and 
interesting phenomena. 

The superexchange which involves the orbital degrees of freedom is 
described by the so-called spin-orbital models \cite{Ole00}, and is 
typically {\it highly frustrated\/} even on a cubic lattice \cite{Fei97}. 
Although this frustration might even lead to the collapse of magnetic (or 
orbital) long-range order in the limit of weak $\propto J_H$, in real 
$e_g$ systems it is largely suppressed by $J_H/U\simeq 0.12$ \cite{Miz96}, 
where $U$ is the intraorbital interaction, and structural phase 
transitions stabilize a particular ordering of occupied orbitals, 
supporting the $A$-type antiferromagnetic (AF) order. Here we show that 
this happens even in the absence of the Jahn-Teller (JT) effect in the 
$e_g$ systems with degenerate orbitals filled either by one hole 
(KCuF$_3$) \cite{Ofz00}, or by one electron (LaMnO$_3$) \cite{Fei99}. 

The transition metal oxides with partly filled $t_{2g}$ orbitals are even
more fascinating. The quantum phenomena are here more important and 
stabilize the {\it coherent orbital liquid\/} ground state in the spin 
$S=1/2$ Mott-insulator LaTiO$_3$ \cite{Kha00}, which preserves the cubic 
symmetry and explains the observed isotropic $G$-type AF order 
\cite{Kei00}. In vanadium compounds rather involved spin-orbital models,
which describe coexisting AF and FM interaction, were recently introduced 
for LiVO$_2$ \cite{Pen97} and V$_2$O$_3$ \cite{Mil00}. The superexchange 
is again frustrated in cubic systems, and $C$-type of AF order, observed 
both in LaVO$_3$ and $T=0$ \cite{Miy00} and in YVO$_3$ $77<T<114$ K 
\cite{Ren00}, can be explained as supported by quantum one-dimensional 
(Q1D) orbital fluctuations \cite{Kha01}.

\section{ Magnetic and orbital order in cuprates and manganites }
\label{sec:eg}

Conceptually the simplest realistic spin-orbital model can be derived 
for $d^9$ ions interacting on a cubic lattice, as in KCuF$_3$. The 
charge excitations $d_i^9d_j^9\rightleftharpoons d_i^8d_j^{10}$ lead to: 
one high spin $^3A_2$ state, and two low-spin $^1E$ and $^1A_1$ states
\cite{Ofz00}. The energy spectrum in Fig. \ref{fig:ex}(a) is obtained 
from the model Hamiltonian which includes the on-site $U$ and $J_H$ 
interactions for degenerate $d$ orbitals \cite{Ole83}, and reproduces 
the exact spectrum \cite{Gri71}. The superexchange is $\propto J_e=
t_{\sigma}^2/U$, where $t_{\sigma}$ is the largest hopping element 
between two $3z^2-r^2$ orbitals along the $c$ axis (note that this is a 
natural unit for the anisotropic hopping between $e_g$ orbitals 
\cite{Ole00}), and is given by 
\begin{equation}
\label{model9}
{\cal H}(d^9)=J\sum_{\gamma}\sum_{\langle ij\rangle\parallel\gamma}
    \Big[({\vec S}_i\cdot {\vec S}_j+S^2)
    {\hat J}_{ij}^{(\gamma)}(d^9) + {\hat K}_{ij}^{(\gamma)}(d^9)\Big],
\end{equation}
where ${\vec S}_i$ are spin $S=1/2$ operators. The operator expressions:
\begin{eqnarray}
\label{j9}
{\hat J}_{ij}^{(\gamma)}(d^9)&=&
(2+\eta p_2-\eta p_3){\cal P}_{\langle ij\rangle}^{\zeta\zeta}
      -\eta(3p_1-p_2){\cal P}_{\langle ij\rangle}^{\zeta\xi},        \\
\label{k9}
{\hat K}_{ij}^{(\gamma)}(d^9)&=&
-[1+\eta(3p_1+p_2)/2]{\cal P}_{\langle ij\rangle}^{\zeta\xi}
-[1+\eta( p_2-p_3)/2]{\cal P}_{\langle ij\rangle}^{\zeta\zeta},
\end{eqnarray}
describe spins and orbital superexchange, with $\eta=J_H/U$, 
$p_1=1/(1-3\eta)$, $p_2=1/(1-\eta)$, and $p_3=1/(1+\eta)$. They depend on 
orbital operators:
\begin{eqnarray}
\label{porbit}
{\cal P}_{\langle ij\rangle}^{\zeta\xi}&=&
 (1/2+\tau^{\gamma}_i)(1/2-\tau^{\gamma}_j)
+(1/2-\tau^{\gamma}_i)(1/2+\tau^{\gamma}_j),       \\
{\cal P}_{\langle ij\rangle}^{\zeta\zeta}&=&
2(1/2-\tau^{\gamma}_i)(1/2-\tau^{\gamma}_j),
\end{eqnarray}
which project on the orbital states, being either parallel
to the bond $\langle ij\rangle$ direction on one site
($P_{i\zeta}=1/2-\tau^{\gamma}_i$) and perpendicular on the other
($P_{j  \xi}=1/2+\tau^{\gamma}_j$), or parallel on both sites. They are 
represented by the orbital operators $\tau^{\alpha}_i$ associated with 
the three cubic axes ($\gamma=a$, $b$, or $c$),
\begin{equation}
\label{orbop}
\tau^{a(b)}_i = ( -\sigma^z_i\pm\sqrt{3}\sigma^x_i )/4, \hskip .7cm
\tau^c_i = \sigma^z_i/2,
\end{equation}
where the $\sigma$'s are Pauli matrices acting on:
$|x\rangle ={\left( \begin{array}{c} 1\\ 0\end{array}\right)},\;
 |z\rangle ={\left( \begin{array}{c} 0\\ 1\end{array}\right)}$,
which transform as $|x\rangle \propto x^2-y^2$ and
$|z\rangle \propto (3z^2-r^2)/\sqrt{3}$.

The superexchange in LaMnO$_3$ couples {\it total spins\/} $S=2$ at the 
$d^4$ Mn$^{3+}$ ions and originates from the charge excitations, 
$d_i^4d_j^4\rightleftharpoons d_i^3d_j^5$ \cite{Fei99}. The $e_g$ part, 
following from $d_i^4d_j^4\rightleftharpoons d_i^3(t_{2g}^3)
d_j^5(t_{2g}^3e_g^2)$ processes, involves FM terms due to the high-spin
$^6A_1$ state, and AF terms due to the low-spin states: $^4A_1$, $^4E$, 
and $^4A_2$ [Fig. \ref{fig:ex}(a)], is orbital dependent. By contrast, 
the $t_{2g}$ part $\propto j_t\simeq 0.09$, which follows from 
$d_i^4d_j^4\rightleftharpoons d_i^3(t_{2g}^3)d_j^5(t_{2g}^4e_g)$ 
excitations, is purely AF and orbital independent. Both terms give
\begin{equation}
\label{model4}
{\cal H}(d^4)=J_e\sum_{\gamma}\sum_{\langle ij\rangle\parallel\gamma}
    \Big[({\vec S}_i\cdot {\vec S}_j+4)
    {\hat J}_{ij}^{(\gamma)}(d^4) + {\hat K}_{ij}^{(\gamma)}(d^4)\Big],
\end{equation}
where the exchange interactions depend on the multiplet structure,
\begin{equation}
{\hat J}_{ij}^{(\gamma)}(d^4)=2\Big[1-\frac{4}{3}\eta(q_3+2q_4)\Big]
    {\cal P}_{\langle ij\rangle}^{\zeta\zeta}
-\frac{2}{15}\eta\Big(36q_1+9q_2+20q_3\Big)
    {\cal P}_{\langle ij\rangle}^{\zeta\xi}+j_t,
\label{j4}
\end{equation}
with $q_1=1/(1-3\eta)$, $q_2=1/(1+2\eta)$, $q_3\simeq 1/(1+8\eta/3)$, 
and $q_4\simeq 1/(1+16\eta/3)$ \cite{Gri71}. The orbital part 
${\hat K}_{ij}^{(\gamma)}(d^4)$ is given in Ref. \cite{Fei99}. 

Both $d^9$ model and $d^4$ model at $j_t=0$ describe strongly frustrated 
superexchange in the limit of $J_H\to 0$, which takes a universal form,
\begin{equation}
\label{h0}
{\cal H}_e^{(0)}=J_e\sum_{\gamma}\sum_{\langle ij\rangle\parallel\gamma}
    \Big[ ({\vec S}_i\cdot {\vec S}_j/S^2+1)
          (1/2-\tau^{\gamma}_i)(1/2-\tau^{\gamma}_j) - 1 \Big].
\end{equation}
Several classical phases have the same energy of $-3J_e$ per site at 
this point \cite{Fei97}: the $G$-AF phases with arbitrary 
occupation of orbitals, and $A$-AF phases with $\langle 
(1/2-\tau^{\gamma}_i)(1/2-\tau^{\gamma}_j)\rangle=0$, as obtained for
staggerred planar orbitals, e.g. for $x^2-y^2/y^2-z^2$. The model 
(\ref{h0}) is qualitatively different from the idealized 
SU(4)-symmetric case \cite{Li98} due to the directionality of $e_g$ 
orbitals.

At finite $J_H$ the degeneracy of classical phases is removed, and the
$A$-AF phase is stable, with two-sublattice alternating orbital order in 
both cuprate (\ref{model9}) and manganite (\ref{model4}) model, 
$|i\mu\sigma\rangle=\cos\theta_i|iz\sigma\rangle\pm\sin\theta_i|ix\sigma
\rangle$, where $\pm$ refers to $i\in A(B)$ sublattice. In the cuprates 
the orbital order given by $\cos 2\theta=(1-\eta/2)/(2+3\eta)$, induces 
FM interactions $J_{ab}$ within the $(a,b)$ planes, and AF interactions 
$J_c$ between them \cite{Ole00}. The AF interactions decrease with 
increasing $J_H/U$ [Fig. \ref{fig:eg}(a)], but still dominate at 
realistic $J_H/U\simeq 0.12$ \cite{Miz96}, explaining why the excitation 
spectra of KCuF$_3$ are dominated by Q1D spin excitations of $S=1/2$ 
spin chains \cite{Ten00}. 

Although the orbital order found in the manganite model (\ref{model4}) 
at $J_H/U=0$ is again $x^2-z^2/y^2-z^2$, and the $A$-AF phase is stable, 
the situation is here {\it qualitatively\/} different as $J_{ab}$ and 
$J_c$ change much faster with increasing $J_H/U$ [Fig. \ref{fig:eg}(b)],
and have similar values in LaMnO$_3$ ($J_H/U\simeq 0.117$ \cite{Miz96}), 
demonstrating the proximity to ferromagnetism which is indeed observed 
in doped manganites \cite{Tok00,Ole00}. Including the (smaller) $t_{2g}$ 
interactions one finds a somewhat enhanced tendency towards 
antiferromagnetism, with the $G$-AF ($A$-AF) phase stable for 
$J_H/U<0.05$ ($J_H/U>0.05$). In order to explain quantitatively the 
experimental ratio $J_c/J_{ab}\simeq 0.7$ in LaMnO$_3$, one has to 
include also the JT effect which stabilizes the orbital order closer to 
$(|x\rangle+|z\rangle)/(|x\rangle-|z\rangle)$ alternation \cite{Fei99}. 
This modification of the orbital ordering changes not only the 
effective magnetic interactions, but also considerably reduces the 
scattering of a hole on spin excitations in LaMnO$_3$ \cite{Bal01}.

\section{ Orbital fluctuations in $t_{2g}$ systems }
\label{sec:t2g}

As in the $d^9$ case, the excitation spectra of $d^2$ and $d^3$ ions in
the $t_{2g}$ subspace \cite{Gri71}, shown in Fig. \ref{fig:ex}(b), may be 
faithfully reproduced with a model Hamiltonian \cite{Ole83} containing 
only two parameters: $U$ and $J_H$, with $J_H$ standing now for the 
Hund's element between two $t_{2g}$ orbitals. As usually, the excitation 
energy to high-spin ($^3A_2$ and $^4A_2$) states is $U-3J_H$, while the 
energy of the next (low-spin) excited states is either $U-J_H$ for $d^2$ 
ions ($^1T_2$, $^1E$), or $U$ for $d^3$ ions ($^2T_1$, $^2E$), 
respectively. The highest excitation energy of $U+2J_H$ is the same for 
$d^2$ ($^1T_1$) and $d^3$ ($^2T_2$) ions.

Each $t_{2g}$ orbital is orthogonal to one of the cubic axes, so we label
them as $a$, $b$ , and $c$ (for instance, $xy$ orbitals are labelled as 
$c$). The superexchange interactions $\propto J=4t^2/U$ follow from the 
hopping between two orbitals {\it active along a given direction 
$\gamma$\/}, for instance between the pairs of $a$ and $b$ orbitals along 
the $c$ axis. Therefore, it is convenient to define pseudospin operators,
${\vec\tau}_i=\{\tau_i^x,\tau_i^y,\tau_i^z\}$, which act in the subspace 
spanned by two active orbital flavors \cite{Kha00,Kha01}. For instance, 
for a bond $\langle ij\rangle\parallel c$, these operators are:
$\tau_i^+=a_i^{\dagger}b_i^{}$, $\tau_i^-=b_i^{\dagger}a_i^{}$, 
$\tau_i^z=\frac{1}{2}(n_{ia}^{}-n_{ib}^{})$, and 
$n_i^{(c)}=n_{ia}^{}+n_{ib}^{}$, where $\{a_i^{\dagger},b_i^{\dagger}\}$
are Schwinger bosons for $a$ and $b$ orbitals. 

The model for titanates follows from the $d_i^1d_j^1\rightleftharpoons 
d_i^0d_j^2$ processes,
\begin{equation}
\label{model1}
{\cal H}(d^1)=J\sum_{\gamma}\sum_{\langle ij\rangle\parallel\gamma}
    \Big[({\vec S}_i\cdot {\vec S}_j+S^2)
    {\hat J}_{ij}^{(\gamma)}(d^1)+{\hat K}_{ij}^{(\gamma)}(d^1)\Big],
\end{equation}
with the exchange constants between $S=1/2$ spins,
\begin{eqnarray}
\label{j1}
{\hat J}_{ij}^{(\gamma)}\!&=&\!
2({\vec\tau}_i\!\cdot\!{\vec\tau}_j\!\!+\!\!n_in_j/4)\!
+\!\eta[(-3r_1\!+\!r_2)(n_{ia}n_{jb}\!+\!n_{ib}n_{ja}\!
+\!n_{ic}\!+\!n_{jc}\!-\!n_{ic}n_{jc})                   \nonumber \\
&+&\!(3r_1\!+\!r_2)(\tau_i^+\tau_j^-\!+\!\tau_i^-\tau_j^+)
 +4(r_2\!-\!r_3)(n_{ia}n_{ja}\!+\!n_{ib}n_{jb})/3]/2,
\end{eqnarray}
depending on: $r_1=1-3\eta$, $r_2=1-\eta$, $r_3=1+2\eta$, while 
${\hat K}_{ij}^{(\gamma)}(d^1)$ stands for purely orbital interactions. 
{\it A priori,\/} the magnetic interactions are anisotropic, and may be 
either AF or FM, depending on the orbital correlations. In the limit of 
$J_H/U=0$ the Hamiltonian (\ref{model1}) takes the form,
\begin{equation}
\label{H0}
{\cal H}^{(0)}=
(J/2)\sum_{\gamma}\sum_{\langle ij\rangle\parallel\gamma}
     [({\vec S}_i\!\cdot\!{\vec S}_j/S^2+1)
      ({\vec\tau}_i\!\cdot\!{\vec\tau}_j+n_in_j/4)-4S/3],
\end{equation}
and shows again a strong frustration of superexchange interactions 
\cite{Kha00}. 
Although formally it resembles the SU(4)-symmetric spin-orbital models 
\cite{Li98} even more than Eq. (\ref{h0}), the pseudospin operators 
${\vec\tau}_i$ have here a different meaning and refer to different 
orbital flavors for each cubic direction $\gamma$. One may also notice 
a certain analogy with the models of valence bond solids \cite{Aff87}, 
but this analogy is again only partial, as the formation of orbital 
singlets in all directions simultaneously is impossible. 

In the mean field approach (MFA) the $G$-AF phase is degenerate with FM 
phases, if $\langle{\vec\tau}_i\!\cdot\!{\vec\tau}_j+\frac{1}{4}n_in_j
\rangle=0$, as realized for alternating orbitals (e.g. for staggered 
$a/b$ orbitals). Such FM states, with anisotropic exchange constants: 
$J_{Fa}$ and $J_{Fc}$ along $a$ ($b$) and $c$ axis [Fig. 
\ref{fig:t2g}(a)], respectively, would be favored classically at finite 
$J_H$. On the contrary, the quantum fluctuations take over, remove the 
anisotropy, and stabilize the {\it orbital liquid\/} state, if the JT 
interactions are weak \cite{Kha00}. Indeed, the spin wave spectrum of 
LaTiO$_3$ is nearly isotropic \cite{Kei00}, showing that the orbital 
moments of $t_{2g}$ ions are fully quanched \cite{Kha00}. Increasing 
$J_H$ almost does not change the exchange constants $J_{AF}$ evaluated 
using the MFA in this state [Fig. \ref{fig:t2g}(a)].

The superexchange interactions between $S=1$ spins in LaVO$_3$ 
\cite{Kha01},
\begin{equation}
{\cal H}(d^2)=J\sum_{\gamma}\sum_{\langle ij\rangle\parallel\gamma}
    \Big[ ({\vec S}_i\cdot {\vec S}_j+1)
    {\hat J}_{ij}^{(\gamma)}(d^2) + {\hat K}_{ij}^{(\gamma)}(d^2) \Big],
\label{model}
\end{equation}
follow from the $d_i^2d_j^2\rightleftharpoons d_i^1d_j^3$ processes
active on the bonds, with 
%
\begin{equation}
\label{j2}
{\hat J}_{ij}^{(\gamma)}\!\!=\!\!
\frac{1}{2}\!\left[(1\!+\!2\eta R)
\Big(\!{\vec\tau}_i\!\cdot\!{\vec\tau}_j
     \!\!+\!\!\frac{1}{4}n_i^{}n_j^{}\!\Big)      
\!-\!\eta r\Big(\!\tau_i^z\tau_j^z\!\!+\!\!\frac{1}{4}n_i^{}n_j^{}\!\Big)
\!-\!\frac{1}{2}\eta R(n_i\!+\!n_j)\right]^{\!(\gamma)},              
\end{equation}
and the orbital term ${\hat K}_{ij}^{(\gamma)}$ given in Ref. 
\cite{Kha01}. The coefficients $R=1/(1-3\eta)$ and $r=1/(1+2\eta)$ follow 
from the multiplet structure of $d^3$ ions [Fig. \ref{fig:ex}(b)]. In the 
limit of $J_H\to 0$ one finds again the frustrated superexchange  
(\ref{H0}). While the orbital liquid cannot stabilize in this case, 
orbital singlets may form along the $c$ direction when $c$ orbitals have 
condensed ($n_{ic}=1$) and the $a$ and $b$ orbitals fluctuate. This gives 
a {\it novel mechanism of ferromagnetic interactions\/} which operates 
already in the limit of $J_H=0$ \cite{Kha01}.

The exchange constants within $(a,b)$ planes ($J_{ab}$) and along 
$c$ axis ($J_c$):
\begin{eqnarray}
\label{jab}
J_{ab}&=&[1-\eta (R\!+\!r)
 +(1+2\eta R-\eta r)\langle n_{ia}n_{ja}\rangle^{(b)}]/4,      \\ 
\label{jc}
J_c&=&[(1\!+\!2\eta R)
 \langle{\vec\tau}_i\!\cdot\!{\vec\tau}_j+1/4\rangle^{(c)}
-\eta r\langle\tau_i^z\tau_j^z+1/4\rangle^{(c)}-\eta R]/2,                                                   
\end{eqnarray}
are given by orbital correlations. Their values at $\eta=0$ were 
obtained from the Bethe ansatz for a Q1D Heisenberg chain, while the
orbital wave spectrum, 
$\omega_k^{\rm C}=[\Delta^2+R^2(1-\cos^2k)]^{1/2}$, with a gap
$\Delta=\{\eta (R+r)[2R+\eta (R+r)]\}^{1/2}$, was used at finite $J_H$.
As a result, one finds increasing FM ($J_c$) and decreasing AF ($J_{ab}$)
exchange constants with increasing $J_H$ [Fig. \ref{fig:t2g}(b)], and 
both interactions have similar values at $J_H/U\simeq 0.15$ \cite{Miz96}. 

While the cubic structure of LaVO$_3$ is almost undistorted \cite{Miy00},
YVO$_3$ has a distorted structure, and $a$ and $b$ orbitals stagger in 
$(a,b)$ planes and repeated themselves along $c$ axis \cite{Ren00}. Such 
ordering can be promoted by the JT effect term which lowers the energy by 
$-2V$ on the bonds along the $c$ axis when $a$ ($b$) orbitals are 
repeated in $C$-type orbital ordered state \cite{Kha01}. Finite $V>0$ 
lowers the energy of the $G$-phase, but the entropy ${\cal S}$ determined 
by orbital excitations increases faster in the $C$-phase, and thus 
induces a transition from $G$-AF to $C$-AF order around $T^*\simeq 0.8J$ 
(Fig. \ref{fig:f}), reproducing qualitatively the first order transition 
observed in YVO$_3$ \cite{Ren00}.

\section{ Summary and open problems }
\label{sec:summa}

In summary, the transition metal oxides with orbital degrees of freedom
show a very fascinating behavior, with various types of magnetic and
{\it orbital order\/}. While $e_g$ orbitals usually order and explain 
$A$-AF phases, further stabilized by the JT effect, the $t_{2g}$ 
orbitals have a generic tendency towards disorder, which leads to the 
orbital liquid in the isotropic $G$-AF phase in LaTiO$_3$. In cubic 
vanadates the JT interactions compete with the {\it orbital disorder\/}, 
and the Q1D orbital fluctuations stabilize the $C$-AF phase in LaVO$_3$, 
and also in YVO$_3$ at finite temperatures. A better understanding of 
these fluctuations is required to explain quantitatively the observed 
phase transitions and the strong reduction of the magnetic order 
parameter in LaVO$_3$ and YVO$_3$. This problem is as urgent as the 
theoretical understanding of the colossal magnetoresistance in the 
manganites.

It is a pleasure to thank Lou-Fe' Feiner, Peter Horsch, Giniyat 
Khaliullin, and Jan Zaanen for a friendly collaboration on this subject
and for numerous stimulating discussions. 
This work was supported by the Committee of Scientific Research (KBN) 
of Poland, Project No. 2~P03B~055~20.


\eject

\begin{figure}
\caption
{Excitation spectra in cubic transition metal oxides for: 
 (a) $e_g$ systems: Cu$^{3+}$ ($d^8$) and Mn$^{2+}$ ($d^5$) ions;
 (b) $t_{2g}$ systems: Ti$^{2+}$ ($d^2$) and  V$^{2+}$ ($d^3$) ions.}
\label{fig:ex}
\end{figure}

\begin{figure}
\caption
{Exchange constants FM $J_{ab}$ (solid lines) and AF $J_c$ (dashed lines) 
 in $A$-AF phase of $e_g$ systems as functions of $J_H/U$ for: 
 (a) cuprates (KCuF$_3$); 
 (b) manganites (LaMnO$_3$), for: $j_t=0$     (thin lines) 
                              and $j_t=0.09$ (heavy lines).}
\label{fig:eg}
\end{figure}

\begin{figure}
\caption
{Exchange constants as functions of $J_H/U$ for $t_{2g}$ systems: 
 (a) $G$-AF ($J_{AF}$, dashed line) and FM ($J_{Fa}$ and $J_{Fc}$, 
 solid lines) phase in titanates; 
 (b) AF $J_{ab}$ (dashed line) and FM $J_c$ (solid line) $C$-AF phase 
 in vanadates (LaVO$_3$).}
\label{fig:t2g}
\end{figure}

\begin{figure}
\caption
{Free energies $F(T)=\langle {\cal H}(d^2)\rangle-T{\cal S}$ (in units 
 of $J$) of: $G$-AF phase obtained with the JT interaction $V=0.65J$ 
 (solid line), and $C$-AF phase for $\eta=0.05$, 0.10 and 0.15 (dashed 
 lines), as functions of temperature $T/J$ 
 (after Ref. \protect\cite{Kha01}).}
\label{fig:f}
\end{figure}

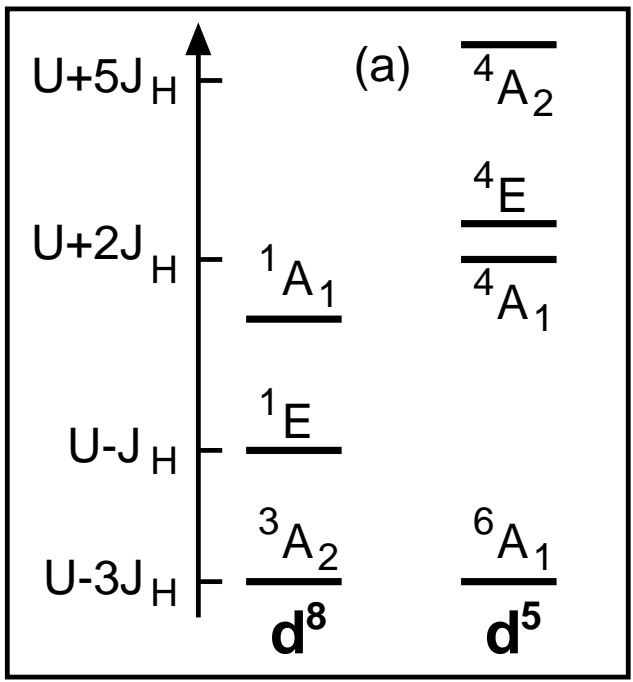
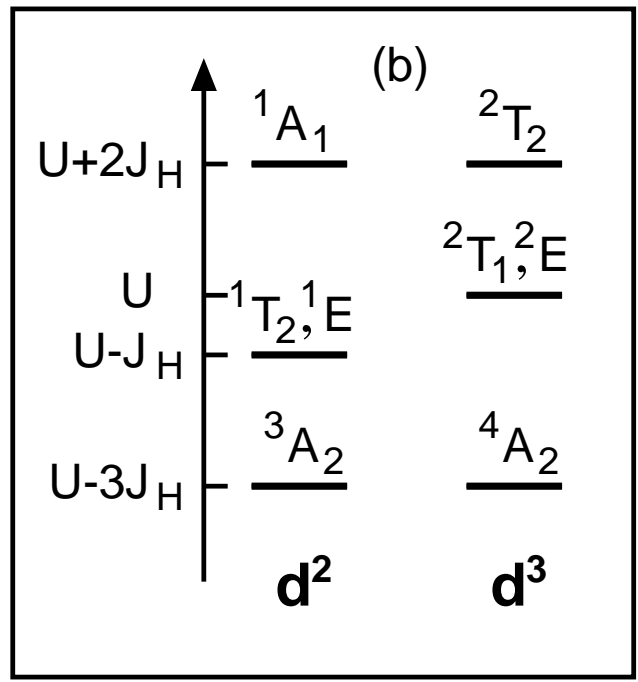
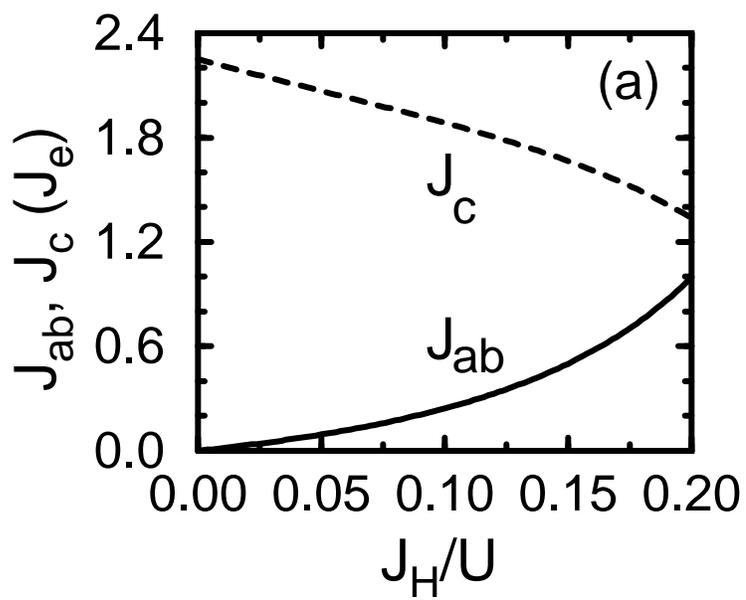
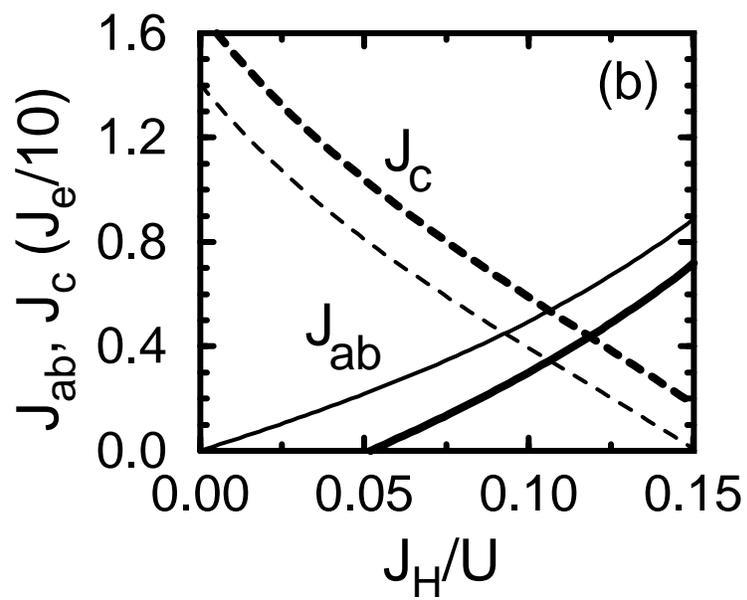
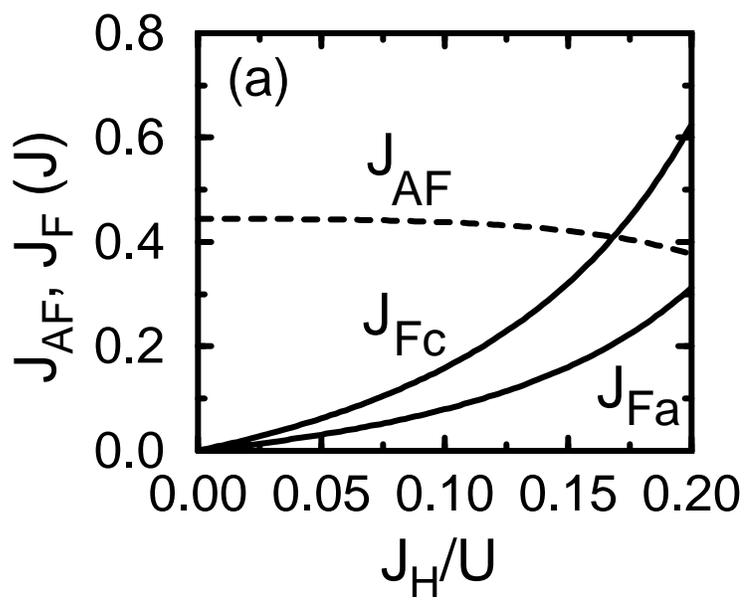
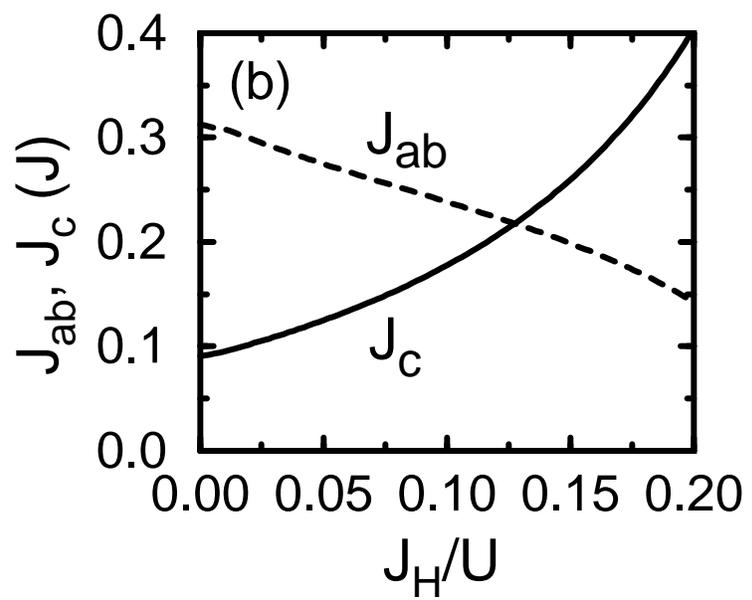
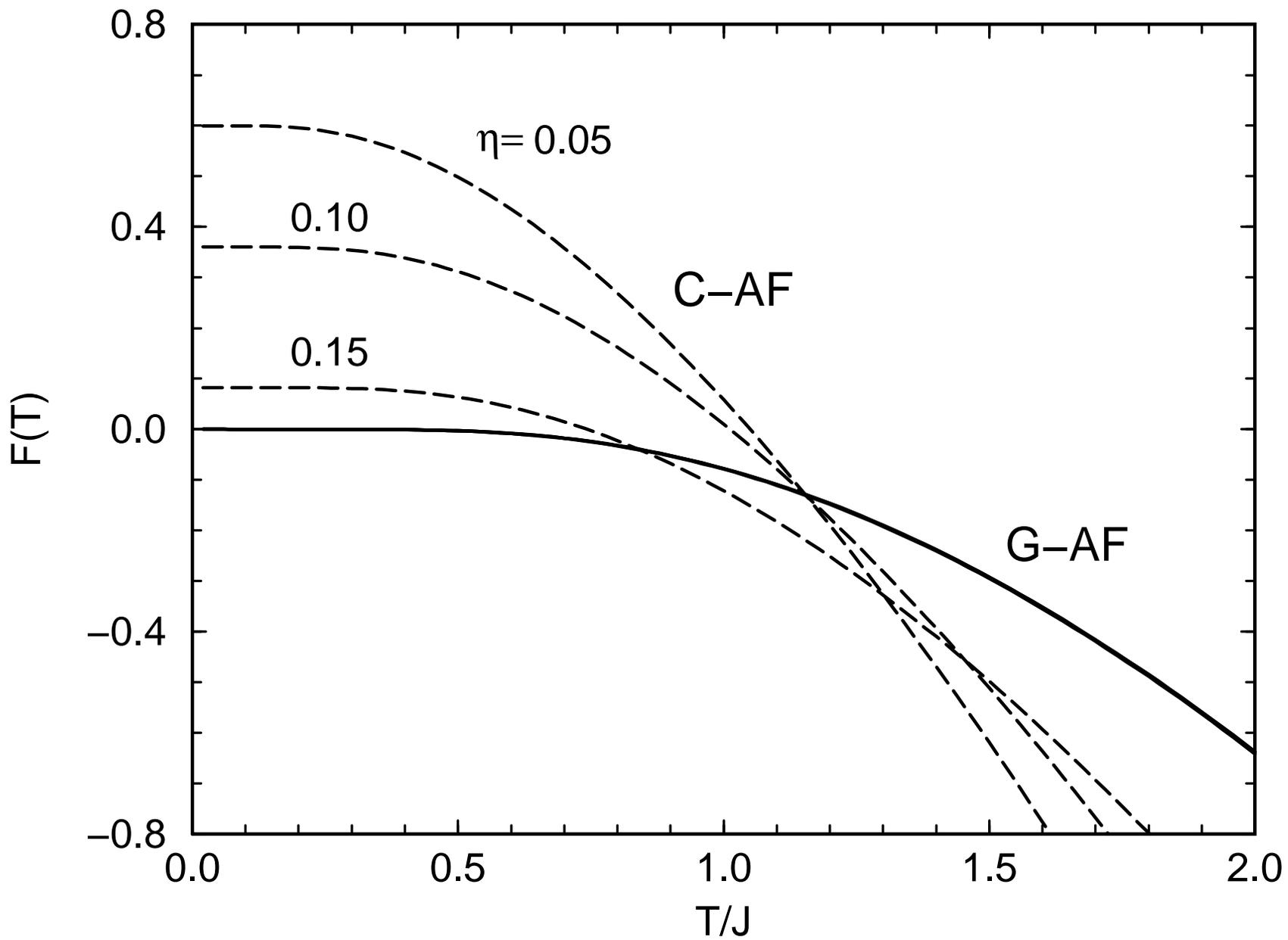
\end{document}